\documentclass[12pt]{article}
\usepackage{amsmath}
\usepackage{amssymb}

\def\sech{{\rm sech}}

\begin{document}

\title{Multipeakons and a theorem of Stieltjes}

\author{R. Beals\thanks{Research supported by NSF grant DMD-9423746}\\
Yale University \\New Haven, CT 06520
\and
D. H. Sattinger\thanks{Research supported by NSF Grant DMS-9501233.}\\
University of Minnesota\\Minneapolis 55455
\and
J. Szmigielski\thanks{Research supported by the Natural Sciences and 
Engineering Research Council of Canada}\\
University of Saskatchewan\\
Saskatoon, Saskatchewan, Canada} 

\maketitle

\centerline{Inverse Problems, {\bf 15}, 1999, Letters L1-L4}

\begin{abstract} A closed form of the multi-peakon solutions of the 
Camassa-Holm equation is found using a theorem of Stieltjes on continued 
fractions.  An explicit formula is obtained for the scattering shifts.
\end{abstract}


\bigskip

Camassa and Holm \cite{ch},\cite{chh} introduced the strongly 
nonlinear equation
\begin{eqnarray}\label{cholm}
 (1-\frac14 D^2)u_t=\frac32 (u^2)_x-\frac18 (u_x^2)_x-
\frac14 (uu_{xx})_x, 
\end{eqnarray}
as a possible model for dispersive waves in shallow water and showed that it 
could formally be integrated using the spectral problem $(D^2+k^2m-1)\psi=0$,
where $m=2(1-\frac14 D^2)u.$ We showed in a previous article 
that a Liouville transformation (\cite{bss1}, equation (5.8))
maps this spectral problem to the string problem
\begin{eqnarray}\label{gstring}
v''(y)+k^2g(y)v(y)=0,
\qquad
-1 <y<1;\qquad v(\pm 1,z)=0,
\end{eqnarray}
where $m(x)=g(y)\,\sech^4(x),$ and $y=\tanh\,x.$ 

The Liouville transformation is independent of $m$, and $m$ may assume 
both positive and negative values. 
The scattering data are invariant under the trans\-for\-ma\-tion,
and consist of a discrete set of eigenvalues $k^2_j$ and associated
coupling constants $c_j$. The evolution of the data under the
flow \eqref{cholm} is $\dot k_j=0$ and $\dot c_j=-2c_j/k_j^{2}$; \cite{bss1}.

The $n$-peakon solution has the form
\begin{eqnarray}\label{npeakon}
u(x,t)=\sum_{j=1}^n p_j(t)\exp(-2|x-x_j(t)|),
\end{eqnarray} 
where the $p_j$, $x_j$ evolve according to a completely integrable  
Hamiltonian system \cite{chh}.  A number of properties of the 
two-peakon and $n$-peakon solutions were obtained from an 
analysis of the dynamical system in \cite{chh}.

In this note we use a continued fraction expansion
and a theorem of Stieltjes to give explicit algebraic 
formulas for the n-peakon solutions.

The multi-peakon solution arises when
$g(y)\,d y$ is a sum of delta functions:  
\begin{eqnarray}\label{gdelta}
g(y)\,{d}y=\sum_{j=1}^ng_j\delta(y-y_j)\,d y,\quad
-1<y_1<\dots <y_n<1.
\end{eqnarray}
Equation \eqref{gstring} is then interpreted in the following sense:
\begin{eqnarray}\label{jump}
v''=0, \quad y_{j-1}<y<y_j; \qquad v'(y_j+)-v'(y_j-)=z\,g_j\,v(y_j),
\end{eqnarray}
for $0\le j\le n$. We let $y_0=-1$ and $y_{n+1}=1$, and we have taken $z=-k^2$ as the spectral parameter.
We denote the lengths of the subintervals by
$l_j=y_{j+1}-y_j,$ for $j=0,\, \dots,\, n$.

Let $\varphi$ and $\psi$ be the solutions of \eqref{jump} that satisfy the 
boundary conditions $\varphi(-1,z)=0,$ $\varphi'(-1,z)=1,$
$\psi(1,z)=0,$ and $\psi'(1,z)=-1$.
The eigenvalues $\{\lambda_j\}$ of \eqref{jump} are the zeros of
$\varphi(1,z)$, which are simple.  The coupling constants $\{c_j\}$ are
characterized by the identities $\varphi(y,\lambda_j)=c_j\psi(y,\lambda_j).$
Differentiating with respect to $y$ and setting $y=1$, 
we find that $c_j=-\varphi'(1,\lambda_j).$

To construct $\varphi$ we define $q_j=\varphi(y_j,z),$ and 
$p_j=\varphi'(y_j-,z)$.  Thus $p_j$ is the value of the derivative
$\varphi'$ in the interval $(y_{j-1},y_j)$.
The $p_j$ and $q_j$ can be obtained recursively from \eqref{jump}
by $q_1=l_0$, $p_1=1$, and
\begin{eqnarray}\label{relation}
q_j-q_{j-1}=l_{j-1}\,p_j,
\qquad
p_j-p_{j-1}=z\,g_{j-1}\,q_{j-1}.
\end{eqnarray}
It follows inductively that $\varphi(1,z)$ is a polynomial of degree
$n$.  

We assume for most of this note that the $n$ point masses $g_j$ are
positive, so that the eigenvalues are negative:
$\lambda_n<\lambda_{n-1}<\dots \lambda_1<0.$
By classical oscillation theory, the $c_j$ alternate signs, with
$c_1>0$.

The scattering data is encoded in the Weyl function
$$
w(z)=\varphi'(1,z)/\varphi(1,z).
$$
Since $\varphi(y,0)=1+y$,  $\varphi(1,0)=2$; and therefore $\varphi(1,z)=2\prod^n_{j=1}(1-z/\lambda_j)$.
Combining these results, we obtain
\begin{eqnarray}\label{partial}
\frac{w(z)}{z}=\frac1{2z}+
\sum^n_{j=1}\frac{a_j}{z-\lambda_j},
\quad a_j=\frac12 c_j\prod_{k\ne j}
(1-\lambda_j/\lambda_k)^{-1}.
\end{eqnarray}
The residues $a_j$ of $w(z)/z$ at $z=\lambda_j$ are positive, since 
both the coupling constants $c_j$
and the successive products alternate signs.

Following Krein \cite{krein2}, we may write the Weyl function as a 
continued fraction
\begin{eqnarray}\label{confrac}
\cfrac{1}{l_n+\cfrac{1}{zg_n+\cfrac{1}{l_{n-1}+\dots}}}.
\end{eqnarray}
This follows by induction on \eqref{relation}.
We have $p_1=1,$ $q_1=l_0,$ $q_2=q_1+l_1p_2,$ $p_2=p_1+zg_1l_0$;
hence
\begin{eqnarray}\nonumber
\frac{p_2}{q_2}=\frac{p_2}{l_0+p_2l_1}=\dots 
=\cfrac{1}{l_1+\cfrac{1}{zg_1+\cfrac{1}{l_0}}},
\end{eqnarray}
Assuming \eqref{confrac} for $n-1$ masses, and adding an additional mass at
$y_n\in(y_{n-1},1)$, we have $p_{n+1}=p_n+zg_nq_n$, 
$q_{n+1}=q_n+l_np_{n+1}$; and
\begin{eqnarray}\nonumber
\frac{p_{n+1}}{q_{n+1}}=\frac{  p_{n+1}  }{   q_n+l_np_{n+1}  }=
\dots
=\cfrac{1}{l_n+\cfrac{1}{zg_n+\cfrac{p_n}{q_n}  }  }
\end{eqnarray}

A classical result of Stieltjes \cite{stieltjes1}
recovers the coefficients of the
continued fraction \eqref{confrac} from the Laurent expansion of $w(z)/z$ at
infinity, obtained from \eqref{partial} by expanding
each $(z-\lambda_j)^{-1}$:
\begin{eqnarray}\label{Ak}
\frac{w(z)}{z}=\sum_{k=0}^\infty \frac{(-1)^kA_k}{z^{k+1}},
\qquad
A_k=\begin{cases}\frac{1}{2}+\sum_{j=1}^n a_j & k=0;\\
\sum_{j=1}^n(-1)^k\lambda_j^ka_j,& k\ge 1.
\end{cases}
\end{eqnarray}
Since the eigenvalues are negative and the $a_j$ are positive, 
each $A_k$ is positive. Stieltjes showed that
such a Laurent series
can be uniquely developed in a continued fraction
\begin{eqnarray}\nonumber
\cfrac{1}{b_1z+ \cfrac{1}{b_2+\cfrac{1}{b_3z+\dots}}},\qquad
b_{2k}=\frac{(\Delta_k^0)^2}{\Delta_k^1\Delta_{k-1}^1},
\quad
b_{2k+1}=\frac{(\Delta_k^1)^2}{\Delta_k^0\Delta_{k+1}^0}.
\end{eqnarray}
Here $\Delta^0_0=1=\Delta^1_0$ and 
$\Delta^0_k$, $\Delta_k^1$, $k\ge 1$, are
the $k\times k$ minors of the Hankel matrix 
\begin{eqnarray}\nonumber
H=\begin{pmatrix} A_0 & A_1 & A_2  & \dots \\
A_1 & A_2 & A_3  & \dots \\
A_2 & A_3 & A_4  & \dots \\
\vdots & & &  \end{pmatrix}.
\end{eqnarray}
whose upper left hand entries are, respectively,  $A_0$ and $A_1$.

Comparing this continued fraction with that for the Weyl function, we obtain
\begin{eqnarray}\label{lj}
l_j=\frac{(\Delta_{n-j}^1)^2}{\Delta_{n-j}^0\Delta_{n-j+1}^0}, \qquad
g_j=\frac{(\Delta_{n-j+1}^0)^2}{\Delta_{n-j+1}^1\Delta_{n-j}^1}.
\end{eqnarray}
Because the $a_j$ are positive, the $\Delta^0_k$ are positive, $k\le n+1$,
and the $\Delta^1_k$ are positive, $k\le n$.  
With $n$ eigenvalues and coupling coefficients, 
 $\Delta^0_{n+2}$ vanishes and the continued fraction 
terminates.
The time dependence of $l_j$ and $g_j$ under \eqref{cholm} is 
determined explicitly from the evolution of the coupling coefficients, 
\eqref{partial}, and \eqref{Ak}.

The multi-peakon solution is given by 
\begin{eqnarray}\nonumber
u(x,t)=\frac12\int_{-\infty}^\infty \exp(-2|x-x'|){d}M(x',t),
\end{eqnarray}
where ${d}M=g(y)({d}y/{d}x){d}x$ is the pull-back of $g\,{d}y$ 
under the transformation $y=\tanh x$.
Hence
\begin{eqnarray}\nonumber
{d}M(x,t)=
\sum_{j=1}^n g_j(t)\,\sech^2(x_j(t))\,
\delta(x-x_j(t))\,dx.
\end{eqnarray}

The positions $x_j(t)$ are obtained recursively from
\begin{eqnarray}\nonumber
x_j=\frac12\log\frac{1+y_j}{1-y_j},
\qquad
y_j=\sum_{k<j}l_k-1.
\end{eqnarray}
Since $\sum l_k=2$ we obtain
\begin{eqnarray}\label{positions}
x_j=\frac12\log\frac{\Lambda_j^-}{\Lambda_j^+},
\qquad
\Lambda_j^-=\sum_{k<j}l_k,
\quad
\Lambda_j^+=\sum_{k\ge j}l_k.
\end{eqnarray}
The n-peakon solution is thus given by \eqref{npeakon} with 
$p_j=\tfrac12 g_j\Lambda_j^-\Lambda_j^+$.

These formulas imply that the asymptotic positions of the peaks are 
given by 
\begin{eqnarray}
x_{n-j+1}\sim t/\lambda_j+\frac12\log 2a_j(0)+\sum_{k<j}\log \Big(
\frac{\lambda_j}{\lambda_k}-1\Big), \quad t\to-\infty; \\
x_{j}\sim t/\lambda_j+\frac12\log 2 a_j(0)+
\sum_{k>j}\log\Big(1-\frac{\lambda_j}{\lambda_k}\Big)
\quad t\to \infty.
\end{eqnarray}
{}From these formulas it can be determined that the phase shift in the peakon  
with speed $1/\lambda_j$ as $t\to\pm\infty$ is
\begin{eqnarray}\label{shift}
\sum_{k>j}\log\Big( 1-\frac{\lambda_j}{\lambda_k}\Big)-\sum_{k<j}
\log\Big(\frac{\lambda_j}{\lambda_k}-1\Big).
\end{eqnarray}
Moreover, the asymptotic height of this same peakon is $-1/\lambda_j$. From this we may conclude that the $n$-peakon solution is asymptotically a sum of single peakons moving with speeds $-1/\lambda_j$.

The two-peakon solution has the form \eqref{npeakon} with 
\begin{eqnarray}
p_1(t)=-\frac{\lambda_1^2a_1+\lambda_2^2a_2}{\lambda_1\lambda_2(\lambda_1a_1+\lambda_2a_2)},
\quad  p_2(t)= -\frac{a_1+a_2}{\lambda_1a_1+\lambda_2a_2};\\[2mm]
x_1(t)=\frac12\log\displaystyle\frac{2(\lambda_1-\lambda_2)^2a_1a_2}
{\lambda_1^2 a_1+\lambda_2^2a_2},\quad
x_2(t)=\frac12\log 2(a_1+a_2),
\end{eqnarray}
where $a_j(t)=a_j(0)e^{2t/\lambda_j}$.
The explicit form for the relative positions and momenta
$x_2-x_1$ and $p_2-p_1$ was given in \cite{chh}.

We have concentrated here on the multi-peakon case, for which the
masses $g_j$ all have the same 
sign and the solutions do not develop singularities.
The procedure is purely algebraic, however, so the formulas hold also for
mixed positive and negative masses: the peakon-antipeakon case.
In this case singularities can occur.

\bigskip

{\bf Acknowledgement} The authors thank H P McKean for some helpful discussions and comments, particularly with regard to the issue of the positivity of $m$.

\end{document}